# FROM NONLINEAR STATISTICAL MECHANICS TO NONLINEAR QUANTUM MECHANICS - CONCEPTS AND APPLICATIONS[*]

Constantino Tsallis

Centro Brasileiro de Pesquisas Físicas and

National Institute of Science and Technology for Complex Systems, Rua Xavier Sigaud 150, 22290-180 Rio de Janeiro-RJ, Brazil

and

Santa Fe Institute, 1399 Hyde Park Road, Santa Fe, NM 87501, USA

We briefly review a perspective along which the Boltzmann-Gibbs statistical mechanics, the strongly chaotic dynamical systems, and the Schroedinger, Klein-Gordon and Dirac partial differential equations are seen as *linear* physics, and are characterized by an index $q = 1$. We exhibit in what sense $q \neq 1$ yields *nonlinear* physics, which turn out to be quite rich and directly related to what is nowadays referred to as *complexity*, or *complex systems*. We first discuss a few central points like the distinction between additivity and extensivity, and the Central Limit Theorem as well as the large-deviation theory. Then we comment the case of gravitation (which within the present context corresponds to $q \neq 1$, and to similar nonlinear approaches), with special focus onto the entropy of black holes. Finally we briefly focus on recent nonlinear generalizations of the Schroedinger, Klein-Gordon and Dirac equations, and mention various illustrative predictions, verifications and applications within physics (in both low- and high-energy regimes) as well as out of it.

## 1 Introduction

The expression *nonlinear physics*, and even *nonlinear science*, has gradually entered in the specialized literature since already a few decades. In particular, it is quite frequently used in connection with complexity in natural, artificial and social sciences. The specific meaning of *linearity* (and concomitantly that of *nonlinearity*) is quite clear when we think of say Maxwell equations or Schroedinger equation, since they are linear in their respective fields. It is however less clear when we think of say statistical mechanics and its dynamical foundations. Here let us start by adopting a very simple and specific sense for the word *linear*, which will nevertheless prove to be, in fact, amazingly powerful. Consider the following (ordinary) differential equation:

$$(1) \qquad \frac{dx}{dy} = ay^q \quad (q \in \mathcal{R}; \, y(0) = 1).$$

This differential equation is *linear* if $q = 1$, and *nonlinear* otherwise. Its solution is given by

$$(2) \qquad y(x) = [1 + (1-q)ax]^{\frac{1}{1-q}} = e_q^{ax} \qquad (e_1^{ax} = e^{ax}).$$

The function $e_q^x$ will from now on be referred to as the *q-exponential* function; its inverse is given by the *q-logarithm* function

$$(3) \qquad \ln_q z = \frac{z^{1-q} - 1}{1 - q} \ (\ln_1 z = \ln z).$$

If we identify $(x, y, a)$ with $(t, \xi, \lambda_q)$ where $t$ is time, $\xi$ is the sensitivity to the initial conditions (defined as $\xi = \lim_{\Delta x(0) \to 0} \frac{\Delta x(t)}{\Delta x(0)}$, $\Delta x(t)$ being the discrepancy of initially close trajectories in a one-dimensional nonlinear dynamical system), and $\lambda_q$ is the generalized Lyapunov coefficient ($\lambda_1$ being the standard Lyapunov exponent), we obtain $\xi = e_q^{\lambda_q t}$. For $q = 1$, this expression yields the standard exponential divergence behavior corresponding to strongly chaotic systems (hence $\lambda_1 > 0$); for $q < 1$ we obtain the power-law behavior $\xi(t) \sim t^{1/(1-q)}$, which is typical of a wide class of weakly chaotic systems (with $\lambda_1 = 0$ and $\lambda_q > 0$).

If instead we identify $(x, y, a)$ with $(E_i, p_i Z, -\beta_q)$, where $E_i$ is the energy of the $i$-th state of a $N$-body Hamiltonian system, $p_i$ is the probability of such state to occur in a stationary state at inverse (effective) temperature $\beta_q = 1/k_B T$, and $Z$ is the partition function, we obtain $p_i Z = e_q^{-\beta_q E_i}$. For $q = 1$, this expression recovers the celebrated Boltzmann-Gibbs (BG) weight for systems at thermal equilibrium; for $q \neq 1$ we obtain the distribution corresponding to stationary or quasi-stationary nonequilibrium states described by nonextensive statistical mechanics [1, 2, 3], i.e., for a wide class of systems whose geometrical-dynamical properties yield pathologies such as lack of ergodicity.

Finally, if we identify $(x, y, a)$ with $(t, \Omega, -1/\tau_q)$, where $\Omega$ is some relaxing relevant quantity defined through

$$(4) \qquad \Omega(t) = \frac{O(t) - O(\infty)}{O(0) - O(\infty)},$$

$O$ being some dynamical observable essentially related to the evolution of the system in phase space (e.g., the time evolution of entropy while the system approaches a stationary state), and $\tau_q$ is a characteristic relaxation time, we obtain $\Omega(t) = e_q^{-t/\tau_q}$. For $q = 1$ we recover the ubiquitous relaxation exponential behavior; for $q > 1$ we recover the power-law $\Omega(t) \sim t^{-1/(q-1)}$ which very many complex systems exhibit.

Let us stress at this point that the above $q$-indices, respectively associated with the sensitivity to the initial conditions ($q_{\text{sensitivity}}$), with the energy distribution at the stationary state ($q_{\text{stationary state}}$), and with the relaxation ($q_{\text{relaxation}}$), typically do *not* coincide among them (typically they satisfy $q_{\text{sensitivity}} \leq 1 \leq q_{\text{stationary state}} \leq q_{\text{relaxation}}$). But they simultaneously become $q_{\text{sensitivity}} = q_{\text{stationary state}} = q_{\text{relaxation}} = 1$ for the linear limit. Arguments exist (see, for instance, [3,4]) which suggest that, for a typical universality class of complex (nonlinear) systems, an infinite number of different $q$-indices are to be associated with various types of physical properties (low- and high-order space-time correlations of various micro-, meso- and macroscopic variables). However, only one or few of them are expected to be independent, all the others being (simple or nontrivial) functions of those few independent, which are dictated by the specific class of systems. In the linear limit, all those indices are frequently expected to merge onto the value $q = 1$, thus recovering the exponential behaviors that are standard for various relevant physical variables. The three $q$-indices mentioned here are frequently referred in the literature as the $q$-triplet, and have been exhibited in the solar wind, the ozone layer around the Earth, the edge of chaos of one-dimensional dissipative maps, and elsewhere [3].

In Section II we focus on the difference between additivity and extensivity for quantities such as the entropy. In Section III we briefly review the $q$-generalizations of the Central Limit Theorem (CLT) and of the theory of large deviations. In Section IV we analyze the case of the black-hole entropy with regard to the (frequently mentioned in the literature) bizarre violation of thermodynamical extensivity. In Section V, we flash some predictions, verifications and applications of the present ideas in natural, artificial and social systems, very specifically in recent high-energy experiments at LHC-CERN and RHIC-Brookhaven.

## 2 Additivity versus Extensivity

Following Penrose [5], we will say that an entropic functional $S(\{p_i\})$ is *additive* if, for two probabilistically independent arbitrary systems $A$ and $B$, it satisfies

$$(5) \qquad S(A + B) = S(A) + S(B).$$

We shall focus here on the entropy [1]

$$(6) \qquad S_q = k_B \sum_{i=1}^{W} p_i \ln_q \frac{1}{p_i} = k_B \frac{1 - \sum_{i=1}^{W} p_i^q}{q - 1} \qquad \left( \sum_{i=1}^{W} p_i = 1 \right),$$

where $W$ is the total number of possible configurations of the system. $S_q$ is the basis of a generalization of the BG statistical mechanics, currently referred (for reasons that will soon become clear) in the literature as *nonextensive statistical mechanics* [1, 6, 7].

The hypothesis $p_{i,j}^{A+B} = p_i^A p_j^B$ straightforwardly implies that

$$(7) \qquad \frac{S_q(A + B)}{k_B} = \frac{S_q(A)}{k_B} + \frac{S_q(B)}{k_B} + (1 - q) \frac{S_q(A)}{k_B} \frac{S_q(B)}{k_B}.$$

Therefore $S_q$ is additive for $q = 1$ (i.e., for $S_1 = S_{BG} = -k_B \sum_{i=1}^{W} p_i \ln p_i$), and nonadditive for $q \neq 1$.

The definition of extensivity is quite different, namely the entropy of a given system is *extensive* if, in the $N \to \infty$ limit, $S(N) \propto N$, where $N$ is the number of elements of the system. Consequently, the additivity only depends on the functional relation between the entropy and the probabilities, whereas extensivity depends not only on that, but also on the nature of the correlations between the elements of the system. Hence, checking the entropic additivity is quite trivial, whereas checking its extensivity for a specific system can be quite hard.

To simply illustrate these features let us consider two deeply different equal-probability situations. If the system is such that, for $N \to \infty$, $W(N) \propto \mu^N$ with $\mu > 1$, we have that the additive BG entropy $S_{BG}(N) = k_B \ln W(N) \propto N$, hence it also is extensive. But if $W(N) \propto N^\rho$ with $\rho > 0$, we have that $S_{BG}(N) \propto \ln N$, i.e., it is nonextensive, whereas the (nonadditive) entropy $S_q(N) = k_B \ln_q W(N) \propto N$ for $q = 1 - 1/\rho$, i.e., it is extensive. Another example with strong correlations between the $N$ elements of the system can be seen in [4]. In this example, once again the *nonadditive* entropy $S_q$ is *extensive* for a special value of $q < 1$, whereas the *additive* entropy $S_{BG}$ is *nonextensive*. Nontrivial physical such examples (more precisely, strongly quantum-entangled magnetic systems) can be found in [8, 9]. Summarizing, to satisfy thermodynamic extensivity of the entropy we must use the BG entropy for systems whose elements are independent or closely so, and we must generically use nonadditive entropies ($S_q$ for specific values of $q$, or even other entropies [12,13]; see also [14]) if the elements are strongly correlated.

# 3 Central Limit Theorems and Large-deviation Theory

If we sum many ($N$ with $N \to \infty$) random variables characterized by the same probability distribution, we obtain (after appropriate centering and scaling) an attractor in the probability space if the variables are (strictly or nearly) independent. This attractor is a Gaussian if the variance (as well as higher-order moments) of the distribution is *finite*, and is a Lévy distribution (also referred to as $\alpha$-stable distribution) if the variance *diverges* (and the distribution asymptotically decays as a power-law).

These two well known theorems of theory of probabilities have been recently $q$-generalized. More precisely, if the random variables that are being summed are strongly correlated in a specific manner (named $q$-independence), then the attractor is a $q$-Gaussian (see hereafter for its definition) if a certain $q$-generalized variance is finite [10], and it is a so called $(q, \alpha)$-stable distribution if that same $q$-generalized variance diverges [11]. The definition of $q$-independence is based on a $q$-generalization of the Fourier transform, which turns out to be a nonlinear integral transform. The $q$-generalization of the inverse Fourier transform exhibits in fact properties that are both delicate and interesting [15, 16, 17, 18].

The $q$-Gaussian distributions straightforwardly emerge from the extremization of the entropy $S_q$ in its continuous form, and are defined as follows:

$$(8) \qquad p_q(x) = \frac{e_q^{-\beta x^2}}{\int dy\, e_q^{-\beta y^2}} \propto \frac{1}{[1 + (q-1)\beta x^2]^{1/(q-1)}} \quad (\beta > 0; q < 3).$$

$q$-Gaussians recover Gaussians for $q = 1$, have a finite support for $q < 1$, and an infinite support for $q \geq 1$; they are normalizable for $q < 3$, have a finite variance for $q < 5/3$ and a diverging one for $5/3 \leq q < 3$. For $q = 2$ they recover the celebrated Cauchy-Lorentz distribution.

Since $q$-Gaussians are attractors in the presence of strong correlations ($q$-independence; see also [19, 20]), they are expected to emerge very frequently in nature. We shall present in Section V several such examples.

Another mathematical pillar of BG statistical mechanics is the theory of large deviations [21]. It consists in the fact that the probability of deviations around the mean value exponentially depends on the number $N$ of independent (or nearly so) realizations ($N \to \infty$), the so called *rate function* of the exponent being related to the BG entropy. An illustration has been recently advanced [22] which suggests that in the presence of strongly correlated realizations (of the $q$-independence type), that same probability behaves $q$-exponentially instead of exponentially, the rate function possibly being related to $S_q$ (see also [23]).

# 4 Reconciling The Black Hole Entropy With Thermodynamics

To be self-contained, let us reproduce here parts of the discussion presented in [24]. In his 1902 book *Elementary Principles in Statistical Mechanics* [25], Gibbs emphatically points that systems involving long-range interactions are intractable within the Boltzmann-Gibbs (BG) theory, due to the divergence of the partition function. As an illustration of his remark he refers specifically to the case of gravitation. This serious difficulty emerges in fact for any $d$-dimensional classical system including two-body interactions whose potential energy asymptotically decays with distance like $1/r^\alpha$ ($r \to \infty$), with $0 \leq \alpha/d \leq 1$. Indeed, under such conditions the potential is not integrable, i.e., the integral $\int_{\text{constant}}^{\infty} dr\, r^{d-1} r^{-\alpha}$ diverges. From the microscopic (classical) dynamical point of view, this is directly related to the fact that the entire Lyapunov spectrum vanishes in the $N \to \infty$ limit, which typically impeaches ergodicity (see [26, 27, 28] and references therein). This type of difficulty is also present, sometimes in an even more subtle manner, in various quantum systems (the *free* hydrogen atom constitutes, among many others, an elementary such example; indeed its BG partition function diverges due to the accumulation of electronic energy levels just below the ionization energy).

Along closely related lines, since the pioneering works of Bekenstein [29] and Hawking [30, 31], it has become frequent in the literature the (either explicit or tacit) acceptance that the black-hole entropy is anomalous in the sense that it violates thermodynamical extensivity. Indeed we read all the time claims that the entropy (*assumed to be the BG one*) of the black hole is proportional to the area of its boundary instead of being proportional to its volume [32, 33, 34, 35, 36, 37, 38, 39, 40]. Similarly we have the so called *area law* [41], which states that the entropy (*once again assumed to be the BG one, or occasionally the Renyi one*) of a class of quantum-entangled $d$-dimensional systems (with $d > 1$) is proportional to the $d$-dimensional area $A_d = L^{d-1} \propto N^{(d-1)/d}$ instead of being proportional to its $d$-dimensional volume $V_d = L^d \propto N$, i.e., where $N$ is the number of elements of the system and $L$ is a characteristic length ($d = 3$ precisely coincides with the case of the black hole).

Strangely enough, Gibbs's crucial remark and the dramatic theoretical features to which it is related are often overlooked in textbooks. Similarly, the thermodynamical violation related to the area law frequently is, somehow, not taken that seriously. Indeed, the inclination of some authors is to consider that, for such complex systems, the entropy is not expected to satisfy thermodynamics. Physically speaking, we consider such standpoint a quite bizarre one. It is shown in [24] how this difficulty can be overcome. We simply argue that the fact (repeatedly illustrated in various manners for strongly quantum-entangled systems, black holes and, generically speaking, for systems satisfying the above mentioned area law) that the Boltzmann-Gibbs-von Neumann entropy is *not* proportional to $N$ precisely shows that, for such strongly correlated systems, *the entropy is not the BG one (or the Renyi one, which, like the BG one, is additive) but a substantially different (nonadditive) one*.

It is clear that, for $N \gg 1$, $N^\rho$ becomes increasingly smaller than $\mu^N$. A similar situation occurs for

$$(9) \qquad W(N) \propto C \nu^{N^\gamma} \quad (C > 0; \nu > 1; 0 < \gamma < 1),$$

which also becomes increasingly smaller that $\mu^N$ (though increasingly larger than $N^\rho$). The entropy associated with $\gamma \to 1$ is of course $S_{BG}$. What about $0 < \gamma < 1$? The answer is in fact already available in the literature (footnote of page 69 in [3], and also in [24]), namely,

$$(10) \qquad S_\delta = k_B \sum_{i=1}^{W} p_i \left( \ln \frac{1}{p_i} \right)^\delta \quad (\delta > 0).$$

The case $\delta = 1$ recovers $S_{BG}$. This entropy is, like $S_q$ for $q > 0$, concave for $0 < \delta \le (1 + \ln W)$. And, also like $S_q$ for $q \ne 1$, it is nonadditive for $\delta \ne 1$. Indeed, for probabilistically independent systems $A$ and $B$, we verify $S_\delta(A + B) \ne S_\delta(A) + S_\delta(B)$ ($\delta \ne 1$).

For equal probabilities we have

$$(11) \qquad S_\delta = k_B \ln^\delta W,$$

hence, for $\delta > 0$,

$$(12) \qquad \frac{S_\delta(A + B)}{k_B} = \left\{ \left[ \frac{S_\delta(A)}{k_B} \right]^{1/\delta} + \left[ \frac{S_\delta(B)}{k_B} \right]^{1/\delta} \right\}^\delta.$$

It is easily verified that, if $W(N)$ satisfies (9), $S_\delta(N)$ is extensive for $\delta = 1/\gamma$. This is in fact true even if

$$(13) \qquad W(N) \sim \varphi(N)\nu^{N^{\gamma}} \qquad (\nu > 1; 0 < \gamma < 1),$$

$\varphi(N)$ being any function satisfying $\lim_{N\to\infty} \ln \varphi(N)/N^{\gamma} = 0$. We can unify $S_q$ (Eq. (6)) and $S_\delta$ (Eq. (10)) as follows [24]:

$$(14) \qquad S_{q,\delta} = k_B \sum_{i=1}^{W} p_i \left( \ln_q \frac{1}{p_i} \right)^\delta \qquad \left( q \in \mathcal{R}; \delta > 0; \sum_{i=1}^{W} p_i = 1 \right)$$

$S_{q,1}$ and $S_{1,\delta}$ respectively recover $S_q$ and $S_\delta$; $S_{1,1}$ recovers $S_{BG}$. Obviously this entropy is nonadditive unless $(q,\delta) = (1,1)$, and it is expansible, $\forall (q,\delta)$. It is concave for all $q > 0$ and $0 < \delta \le (qW^{q-1}-1)/(q-1)$. In the limit $W \to \infty$, this condition becomes $0 < \delta \le 1/(1-q), \forall q \in (0,1)$, and any $\delta > 0$ for $q \ge 1$.

We can address now the area law. It has been verified for those anomalous $d$-dimensional systems (with $d > 1$) that essentially $\ln W(N) \propto L^{d-1} \propto N^{(d-1)/d}$, which implies that $W(N)$ is of the type indicated in (13) with $\gamma = (d-1)/d$. Therefore, $S_{q,\delta} = S_{1,\delta}$ for $\delta = d/(d-1)$ is extensive, thus satisfying thermodynamics. At the present state of knowledge we cannot exclude the possibility of extensivity of $S_{q,\delta}$ for other special values of $(q,\delta)$, particularly in the limit $\delta \to \infty$. Indeed, assume for instance that we have $\varphi(N) \propto N^\mu$ in (13), and take the limit $\gamma \to 0$, hence $\delta \to \infty$. The condition $\lim_{N\to\infty} \ln \varphi(N)/N^{\gamma} = 0$ is satisfied for any $\gamma > 0$, but it is violated for $\gamma = 0$, which opens the door for $S_q$, or some other nonadditive entropic functional, being the thermodynamically appropriate entropy.

For example, for the $d = 1$ gapless fermionic system in [8], we have analytically proved the extensivity of $S_q$ for a specific value of $q < 1$ which depends on the central charge of the universality class that we are focusing on. For the $d = 2$ gapless bosonic system in [8], we have numerically found that, once again, it is $S_q$ with a value of $q < 1$ the entropy which is extensive and consequently satisfies thermodynamics. This kind of scenario might be present in many $d$-dimensional physical systems for which $\ln W(L) \propto \ln_{2-d} L$ (i.e., $\propto \ln L$ for $d = 1$, and $\propto L^{d-1}$ for $d > 1$) [1].

Summarizing, the thermostatistics of systems or subsystems whose elements are strongly correlated (for instance, through long-range interactions, or through strong quantum entanglement, or both, like possibly in quantum gravitational dense systems or subsystems) should be based on nonadditive entropies such as $S_{q,\delta}$ (Eq. (14)), and typically not on the usual Boltzmann-Gibbs-von Neumann one. An illustration of the type of back-and-forth arguments that are plethorically present in the literature can be seen in [31]. We read in its Abstract (see Fig. 1): *A black hole of a given mass, angular momentum, and charge can have a large number of different unobservable internal configurations which reflect the possible different initial configurations of the matter which collapsed to produce the hole [...]*, and also *This means that the standard statistical-mechanical canonical ensemble cannot be applied when gravitational interactions are important.* In the last of these sentences, Hawking refers to something which is undoubtedly true, and already known by Gibbs himself [25], i.e., that the BG exponential distribution cannot be used. Nevertheless, in the few lines just above, the formula that is adopted for the entropy precisely is the famous BG one, disregarding the crucial fact that that formula is but the equal-probability particular case of the BG entropic functional from which the BG distribution is (straightforwardly) derived!

---

[1] Logarithmic corrections to these asymptotic behaviors are also possible. For example, for a class of (connected bipartitions of) free-fermion gases it has been recently found $\ln W(L) \propto L^{d-1} \ln L$ [42]. We verify that this expression and $\ln_{2-d} L$ coincide for $d = 1$ but involve a logarithmic discrepancy for $d > 1$. For this class of systems we have $W(N) \propto C\nu^{N^{\gamma} \ln N}$ $(C > 0; \nu > 1; 0 < \gamma < 1)$ (or even $W(N) \propto C(N)\nu^{N^{\gamma} \ln N}$, $C(N)$ being a slowly-varying pre-factor) instead of (9).

## 5 illustrative Predictions, Verifications and Applications, Including $q$-generalized Schroedinger, Klein-Gordon and Dirac Equations

Following Eq. (2), the plane wave can be $q$-generalized as follows [43, 44]: $e_q^{i(kx-\omega t)}$. By using this class of functions (normalizable for $1 < q < 3$, in contrast with the well known non-normalizability of standard plane waves) it is possible to generalize into nonlinear forms: (i) The integral representation of the Dirac delta [45, 46, 47, 48], (ii) The Schroedinger equation [49, 50], and its classical field theory [51]; and (iii) The Klein-Gordon and the Dirac equations [49].

Predictions, verifications and applications have been performed in high-energy physics (in the CMS, ALICE and ATLAS Collaborations at the LHC/CERN, the PHENIX Collaboration at RHIC/Brookhaven, AUGER Project, among various others) [52, 53, 54, 55, 56, 57, 58, 59, 60, 61, 62, 63, 64], spin-glasses [65], cold atoms in optical lattices [66], trapped ions [67], anomalous diffusion [68], dusty plasmas [69], solar physics [70, 71, 72, 73, 74, 75, 76], conservative and dissipative many-body systems [26, 27, 28, 77, 78, 79], finance [80], to mention but a few.

## 6 Final remarks

The size of a geometric object such as a line, a plane, a body, a fractal, is efficiently determined (through a number which is *neither zero nor infinity*) by respectively asking the length, the area, the volume, or the measure in its Hausdorff dimension. In other words, it is the geometric nature of the object which determines the useful question to be asked in order to know its size. In complete analogy, the entropic functional to be efficiently used for a class of probabilistic/thermostatistical system is not universal, but it is determined by the nature of the correlations between its elements (in particular, if this correlation is weak or inexistent, we must use the Boltzmann-Gibbs entropy). The basic criterion for choosing the appropriate functional form is to impose that, for that class of systems, it satisfies thermodynamics, i.e., the extensivity of the entropy. All other physical (dynamical, geometric) properties are believed (as exhibited in some particular instances) to consistently follow from this basic choice. We have shown in this brief review how this philosophy can be applied in sensibly different systems, and in particular in black holes. The (practical and epistemological) correctness of this approach is supported (analytically, numerically, experimentally and observationally) by a wide amount of predictions, verifications and applications in natural, artificial and social systems, some of which have been mentioned here. It is clear that such a physical structure must rely on some basic mathematical foundations, such as central limit theorems and related properties. Although quite succinctly, this has also been addressed here. For further information and a regularly updated bibliography the reader is invited to check [81].

## Black holes and thermodynamics*

S. W. Hawking †

*California Institute of Technology, Pasadena, California 91125
and Department of Applied Mathematics and Theoretical Physics, University of Cambridge, Cambridge, England*
(Received 30 June 1975)

A black hole of given mass, angular momentum, and charge can have a large number of different unobservable internal configurations which reflect the possible different initial configurations of the matter which collapsed to produce the hole. The logarithm of this number can be regarded as the entropy of the black hole and is a measure of the amount of information about the initial state which was lost in the formation of the black hole. If one makes the hypothesis that the entropy is finite, one can deduce that the black holes must emit thermal radiation at some nonzero temperature. Conversely, the recently derived quantum-mechanical result that black holes do emit thermal radiation at temperature $\kappa h/2\pi k c$, where $\kappa$ is the surface gravity, enables one to prove that the entropy is finite and is equal to $c^3 A/4 G h$, where $A$ is the surface area of the event horizon or boundary of the black hole. Because black holes have negative specific heat, they cannot be in stable thermal equilibrium except when the additional energy available is less than 1/4 the mass of the black hole. This means that the standard statistical-mechanical canonical ensemble cannot be applied when gravitational interactions are important. Black holes behave in a completely random and time-symmetric way and are indistinguishable, for an external observer, from white holes. The irreversibility that appears in the classical limit is merely a statistical effect.

**Figure 1** Abstract of [31].

## Acknowledgments

I acknowledge particularly fruitful conversations with M.J. Duff, which, during the event at the Vatican, motivated me to explore in more detail the black-hole entropy case. Partial financial support from CNPq, FAPERJ and CAPES (Brazilian agencies) is acknowledged as well.